\documentclass[12pt,preprint]{emulateapj}
\usepackage{graphicx}

\shorttitle{Constraining Orbital Parameters}
\shortauthors{Stephen R. Kane \& Kaspar von Braun}
\slugcomment{Submitted for publication in the Astrophysical Journal}

\begin{document}

\title{Constraining Orbital Parameters Through Planetary Transit
  Monitoring}
\author{Stephen R. Kane, Kaspar von Braun}
\affil{NASA Exoplanet Science Institute, Caltech, MS 100-22, 770
  South Wilson Avenue Pasadena, CA 91125, USA}
\email{skane@ipac.caltech.edu}


\begin{abstract}

The orbital parameters of extra-solar planets have a significant
impact on the probability that the planet will transit the host
star. This was recently demonstrated by the transit detection of HD
17156b whose favourable eccentricity and argument of periastron
dramatically increased its transit likelihood. We present a study
which provides a quantitative analysis of how these two orbital
parameters affect the geometric transit probability as a function of
period. Further, we apply these results to known radial velocity
planets and show that there are unexpectedly high transit
probabilities for planets at relatively long periods. For a
photometric monitoring campaign which aims to determine if the planet
indeed transits, we calculate the expected transiting planet yield and
the significance of a potential null result, as well as the subsequent
constraints that may be applied to orbital parameters.

\end{abstract}



\section{Introduction}\label{introduction}

With the number of known extra-solar planets exceeding 300,
statistical interpretations of the distribution of orbital parameters
are becoming increasingly significant. These parameter distributions
help us unlock the mysteries surrounding the planet formation process
to which many challanges have been presented, not the least of which
contains the mechanisms that drive planetary migration
\citep{arm07}. \citet*{for08b} showed that transit light curves in
particular can be used to characterize orbital eccentricities and
hence give further insight into the global eccentricity distribution.

In terms of the sheer number of transit light curves, the major
contributors have been the shallow wide-field surveys such as the
Transatlantic Exoplanet Survey (TrES) \citep{man07}, the XO project
\citep{joh08}, the Hungarian Automated Telescope Network (HATNet)
\citep{pal08}, and SuperWASP \citep{and08}. In addition, there have
been at least five cases in which planetary transits were detected
through photometric follow-up of planets already known via their
radial velocity (RV) discoveries. These five planets are HD 209458b
\citep{cha00,hen00}, HD 149026b \citep{sat05}, HD 189733b
\citep{bou05}, GJ 436b \citep{gil07}, and HD 17156b
\citep{bar07a}. The case of HD 17156b is of particular interest since
it is a 21.2 day period planet which happens to have a large
eccentricity ($e = 0.67$) and an argument of periastron which places
the periapsis of its orbit in the direction toward the observer and
close to parallel to the line of sight, resulting in an increased
transit probability.

Conversely, the dominant sources of RV planet discoveries have been
the California \& Carnegie Planet Search \citep{mar97} and the High
Accuracy Radial velocity Planet Searcher (HARPS) \citep{pep04}
teams. However, in the near future we can expect to see larger-scale
surveys \citep*{kan07b} and new instruments \citep{li08} which will
increase both the number and diversity of known planets. There have
been suggestions regarding the strategy for photometric follow-up of
these radial velocity planets at predicted transit times
\citep{kan07a} and the instruments that could be used for such surveys
\citep{lop06a}. Some attempts have been made to detect these possible
transits \citep{lop06b,sha06} which have thus far been unsuccessful.

This paper discusses the effect of orbital parameters on the geometric
transit probability of planets. We calculate orbital constraints that
may be applied, particularly in the absence of transit signatures in
photometric follow-up observations. Section 2 describes how the
eccentricity and argument of periastron of known planetary orbits
affect transit probability. It further presents applications of this
effect to known RV planets and discusses how uncertainties in the
orbital parameter values affect the reliability of the ephemeris
calculations. In Section 3, we show how orbital constraints can be
applied in the absence of a photometrically detected transit signal,
and we discuss the potential transit yield and statistical
significance of a scenario in which no transits are found in a large
sample of RV planets. We summarize and conclude in Section 4.


\section{Transit Probability}\label{transit_probability}

Recent work by \citet{bar07b} and \citet{bur08a} showed that higher
eccentricities of planetary orbits will increase their transit
probabilities and, consequently, expected yield for transit
surveys. In this Section, we demonstrate the combined effect of the
eccentricity and argument of periastron on transit probability. For
explanations of the orbital parameters, including the argument of
periastron $\omega$, we refer the reader to \citet{kan07a} and
\citet{bar07b}. We first explicitly derive the dependence of transit
probability $P_t$ as a function of eccentricity $e$, argument of
periastron $\omega$, and orbital semi-major axis $a$ (i.e.,
period). We discuss this dependence of $P_t$ specifically with respect
to $\omega$ and period, apply the results to a sample of 203
exoplanets compiled in \citet{but06}, and briefly discuss how
ephemeris calculations (and thus planning of photometric follow-up
observations) are affected by uncertainties in $e$ and $\omega$.


\subsection{Orbital Configuration}\label{orbital_configuration}

For a circular orbit the geometric transit probability is proportional
to the inverse of the semi-major axis, $a$, such that the inclination
of the planet's orbital plane $i$ must satisfy
\begin{equation}
  a \cos i \leq R_p + R_\star
\end{equation}
where $R_p$ and $R_\star$ are the radii of the planet and star
respectively \citep{bor84}. For an eccentric orbit, the transit
probability, $P_t$, may be expressed as
\begin{equation}
  P_t = \frac{R_p + R_\star}{a (1 - e \cos E)}
  \label{transit_prob1}
\end{equation}
where $e$ is the eccentricity of the orbit and $E$ is the eccentric
anomaly. The eccentric anomaly and the true anomaly, $f$, are related
to each other by
\begin{equation}
  cos E = \frac{e + \cos f}{1 + e \cos f}
  \label{cose}
\end{equation}
where the true anomaly is defined as the angle between the direction
of periapsis and the current position of the planet in the orbit.
Equation \ref{transit_prob1} can then be evaluated at each point in
the planetary orbit. The transit probability can also be described in
terms of the geometry of an ellipse. For an elliptical orbit, the
separation of the planet and star is
\begin{equation}
  r = \frac{a (1 - e^2)}{1 + e \cos f}.
  \label{separation}
\end{equation}
As shown by \citet{kan07a}, the place in a planetary orbit where it is
possible for a transit to occur (where the planet passes the star-observer
plane that is perpendicular to the plane of the planetary orbit) is when
$\omega + f = \pi / 2$. The transit probability may then be re-expressed as
\begin{equation}
  P_t = \frac{(R_p + R_\star)(1 + e \cos (\pi/2 - \omega))}{a (1 - e^2)}
  \label{transit_prob2}
\end{equation}
consistent with the findings of \citet{bar07b}. Equations
\ref{transit_prob1} and \ref{transit_prob2} both yield the same result
based upon the orbital configuration, but Equation \ref{transit_prob2}
clearly shows the major role played by the the values of $e$ and
$\omega$ in determining the likelihood of a planet transiting the
parent star.


\subsection{Argument of Periastron Dependence}\label{aop_dependence}

\begin{figure}
  \includegraphics[angle=270,width=8.2cm]{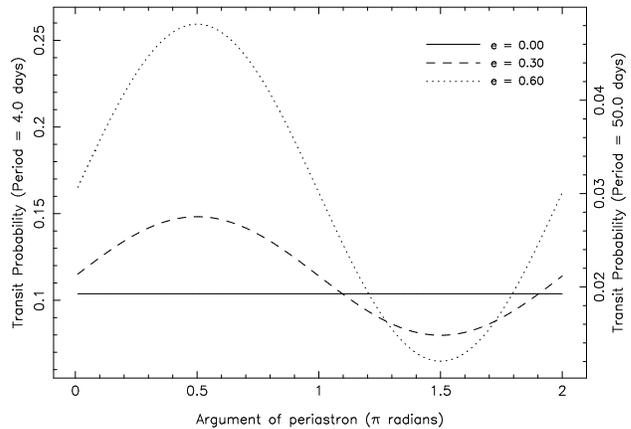}
  \caption{Dependence of geometric transit probability on the argument
    of periastron, $\omega$, for eccentricities of 0.0, 0.3, and 0.6,
    plotted for periods of 4.0 days (left ordinate) and 50.0 days
    (right ordinate). Stellar and planetary radii are assumed to be a
    Jupiter and solar radius, respectively. For details, see \S
    \ref{orbital_configuration} and, in particular, Equation
    \ref{transit_prob2}.}
  \label{fig1}
\end{figure}

Equation \ref{transit_prob2} states the dependence of transit
probability on the argument of periastron. As we rotate the semi-major
axis of the orbit around the star we can observe how the transit
probability varies. This dependence is shown in Figure \ref{fig1} for
eccentricities of 0.3 (dashed line) and 0.6 (dotted line) in
comparison with the constant transit probability for a circular orbit
(solid line). Since the shape of this variation is independent of
period, $P$, the y-axes are scaled for both a 4.0 day and 50.0 day
period orbits. Figure \ref{fig1} assumes a Jupiter radius and a solar
radius for the values of $R_p$ and $R_\star$ respectively. Note that
$P_t$ scales linearly with the sum of these values (Equation
\ref{transit_prob2}).

The peak transit probability occurs at $\omega = \pi / 2$, and the
corresponding increase in $P_t$ as compared to a circular orbit can be
significant: a factor of 1.5 for $e = 0.3$ and a factor of 2.5 for $e
= 0.6$. Moreover, the fraction of the orbital path which produces a
higher value of $P_t$ than the circular orbit with the same period
(corresponding to the fraction of range in $\omega$ for which the
dotted or dashed line is above the solid line in Figure \ref{fig1})
increases with increasing eccentricity.

\begin{figure}
  \includegraphics[width=8.2cm]{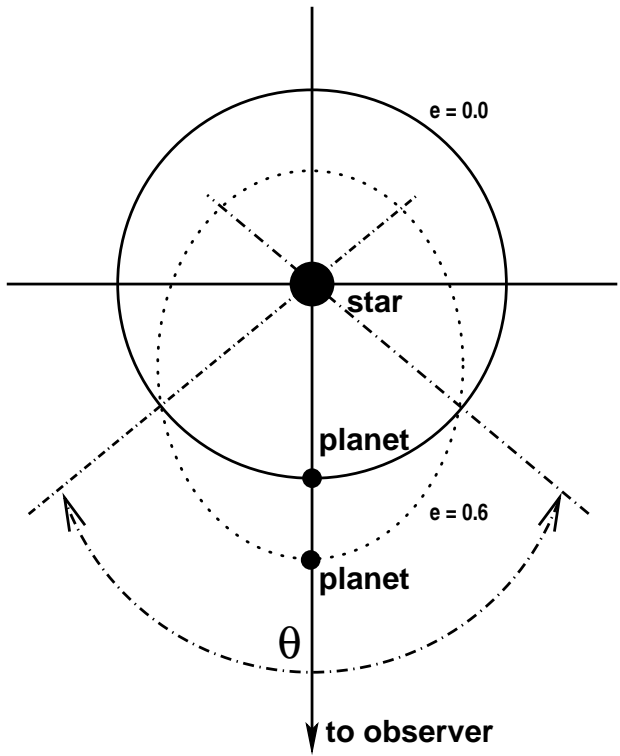}
  \caption{A view from above the orbit pole of a circular (solid line)
    and eccentric ($e = 0.6$; dotted line) planetary orbit for $\omega
    = 3 \pi / 2$. The angle $\theta$ corresponds to the range of
    orbital orientations for which an elliptical orbit has a lower
    transit probability than a circular orbit with the same period
    (see \S \ref{orbital_configuration} and Figure
    \ref{fig1}).}
  \label{fig2}
\end{figure}

The fraction of orbital orientations with $e \neq 0$ producing {\it
  lower} transit probabilities than the corresponding circular orbits
is made clear in Figure \ref{fig2} in which a view from above the
orbit pole of two planetary orbits is depicted. The range of $\omega$
in Figure \ref{fig1} that produces lower values of $P_t$ than a
circular orbit corresponds to the angle between the intersection
points shown in Figure \ref{fig2} for which the planet is located
outside the circular orbit. For an eccentricity of 0.6 this angle is
$\theta = 105\degr$ and decreases with increasing
eccentricity. However, the Keplerian nature of the orbit is such that,
although the larger fraction of the orbital path is spent close to the
star, the larger fraction of time is spent farther away from the star
\citep{bar07b}. This is a crucial aspect in designing a photometric
follow-up campaign to monitor RV planets in eccentric orbits for
possible transits.

\subsection{Period Dependence}\label{period_dependence}

As demonstrated in Figure \ref{fig1}, the peak transit probability
increases with eccentricity. Although the shape of $P_t = f(\omega)$
is independent of period, the magnitude of $P_t$ changes as a function
of period (Figure \ref{fig1}). Consequently, the fractional increase
in $P_t$ for eccentric orbits can be substantial, as shown in \S
\ref{aop_dependence} and argued by \citet{bar07b},

The current distribution of eccentricities for the known extra-solar
planets indicates that orbits within 0.1~AU tend to be forced into
nearly circular orbits through tidal circularization, whereas longer
period orbits can possess a great range of eccentricities
\citep{for08a}. Indeed most of the planets beyond 0.1~AU have
eccentricities in excess of 0.3. Thus, it is the longer-period planets
whose transit probabilities are more likely to be affected by
eccentricities than the short-period ones.

\begin{figure}
  \includegraphics[angle=270,width=8.2cm]{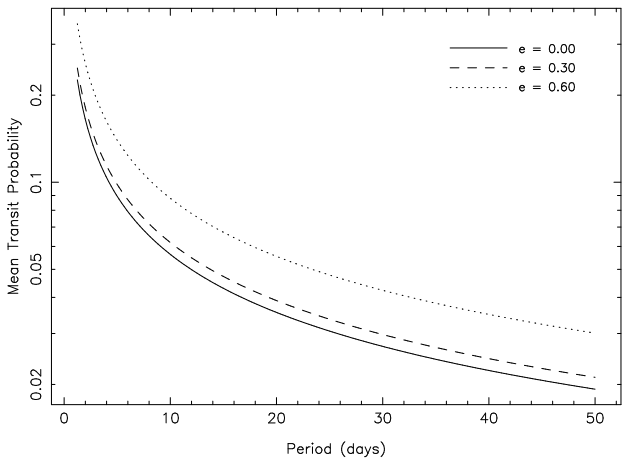}
  \caption{The mean transit probability on a logarithmic scale,
    averaged over all values of $\omega$ (cf. Figure \ref{fig1}), as a
    function of period, for eccentricities of 0.0, 0.3, and
    0.6.}
  \label{fig3}
\end{figure}

In Figure \ref{fig3} we show mean transit probability as a function of
period after averaging over $0 \leq \omega \leq 2 \pi$, for the period
range $1 \leq P \leq 50$ days. Eccentricities of 0.0, 0.3, and 0.6 are
shown with solid, dashed, and dotted lines, respectively. As expected,
we see that doubling the eccentricity from 0.3 to 0.6 creates a
significant increase in the mean transit probability. Most affected
are the longer period planets whose eccentric orbits can raise their
likelihood of transit from a negligible value to a statistically
viable number for photometric follow-up.

\subsection{Application to Known Exoplanets}\label{application}

If we assume circular orbits for each of the known exoplanets, the
transit probability at intermediate to long period orbits makes
photometric searches for planets in those regimes
impractical. However, applying the orbital parameters of $e$ and $\omega$
should in general lead to an overall more favourable
situation for transit detection. Depending on the brightness of the
host star and the cadence of the RV observations, a reasonable
estimate of these two parameters is normally extracted from the RV
fitting.

\begin{figure}
  \includegraphics[angle=270,width=8.2cm]{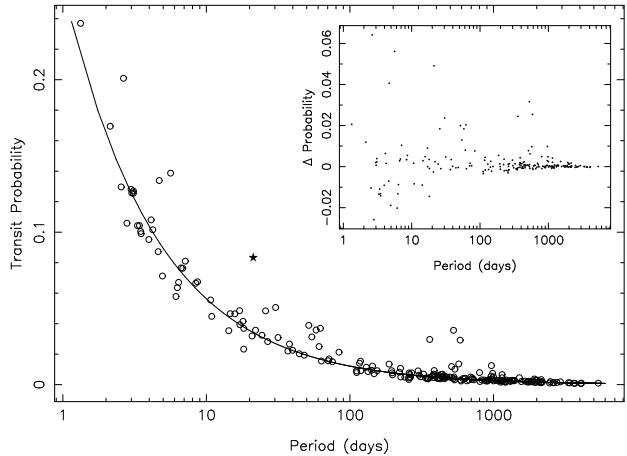}
  \caption{The geometric transit probability for a circular orbit
    (solid curve) along with the transit probability for 203 RV
    planets from \citet{but06} calculated from their orbital
    parameters (open circles). HD 17156b is indicated by a 5-pointed
    star. The sub-panel plots the difference in $P_t$ between the the
    actual orbit and a hypothetical circular orbit for each of the
    planets.}
  \label{fig4}
\end{figure}

Figure \ref{fig4} shows the transit probability calculated from
orbital parameters provided by \citet{but06} for planets with
estimates of $e$ and $\omega$ (203 planets in total). The transit
probabilities are plotted against period, but are calculated from the
semi-major axis, $a$, using Equation \ref{transit_prob2}. For the
purposes of providing an approximate comparison of the relative
transit probabilities, we assume a Jupiter and Solar radius for the
values of $R_p$ and $R_\star$, respectively. Hence, we can include the
transit probability for a circular orbit, shown in the figure as a
solid line.  In addition, the sub-panel in the plot shows the
difference in $P_t$ between the actual orbit and a hypothetical
circular one of the same period (residuals). The mean value of the
residuals for all 203 planets is positive but relatively small ($4.13
\times 10^{-5}$), and is dominated by the low transit probability of
the long period planets. The mean residual of planets with $P < 100.0$
days, however, yields an overall increase of $\sim 0.5$\% in $P_t$.

HD~17156b, a transiting planet with 21.2 day period \citep{bar07a}, is
shown as a 5-pointed star. Its transit probability is greatly
increased by its orbital parameters. Note that the actual $P_t$ of
HD~17156b is larger than the 5\% shown in Figure \ref{fig4} since the
radius of the host star is 1.47~$R_{\sun}$. At longer periods, the
planets with the largest residuals are HD~156846b, HD~4113b, and
HD~20782b, which have periods of 359.51, 526.62, and 585.86 days,
respectively. The probability residuals for these three planets are
0.024, 0.032, and 0.025 respectively, the effect of which is to raise
their transit probabilities to the same level as HD~17156b if it were
in a circular orbit. It is worth noting that these three planets all
have eccentricities close to 0.9 which is undoubtedly the primary
cause of the increased transit probability.

The increased transit probabilities of eccentric planets motivate
photometric follow-up programs of RV planets. Compared to transit
surveys, such programs require much less telescope time since the time
of transit is, in principle, known. However, it was shown by
\citet{kan07a} that reliable constraints on $e$ and $\omega$ are
needed to avoid significant offsets in predicted transit times. This
is particularly true of long period planets.  In the case of planets
HD 156846b, HD 4113b, and HD 20782b, the uncertainties cited in
\citet{but06} indicate that the values of $e$ are all constrained to
$\pm 0.03$ and the values of $\omega$ are constrained to $\pm
3.0$\degr (compared to the mean and median values for all 203 planets
of $\delta \omega = 20\degr$ and 10\degr, respectively). Thus,
ephemerides for these planets can relatively reliably be determined
from RV fit parameters alone. The transit duration is on the order of
12 hours for these planets, ensuring that one will practically never
observe both the ingress and the egress of the transit during a single
orbit from the ground. However, the large transit duration and
relative low uncertainties in $e$ and $\omega$ will increase the
chances of observing at least a partial transit during the predicted
observing window.


\section{Constraining Orbital Parameters}

In \S 2, we discussed transit probability as a function of various
system parameters as well as aspects of potential photometry follow-up
campaigns. Here we focus on what can be learned from the presence and
absence of a planetary transit in follow-up observations. 

For a transiting planet the physical properties (such as the mass, radius, and
density) can be calculated, leading to determination of (as opposed to
constraints on) system parameters of the planet.  Furthermore, the orbital
inclination can be compared with the plane of stellar rotation \citep{win07}
and used to test planetary models regarding co-planar orbits.

However, even the absence of a planetary transit signature in photometric data
can lead to interesting constraints on the orbital parameters. Below, we
elaborate on these constraints and apply the results to the aforementioned
\citet{but06} sample of RV planets.


\subsection{Orbital Radius versus Stellar Radius}
\label{radii}

One implicit assumption in the derivation of the transit probability by
\citet{bor84} is that the planet remains well outside the star in order to
produce the solid angle of the planet's shadow \citep[see also][]{bar07b}. As
a result, the calculation of $P_t$ in Equation \ref{transit_prob2} becomes
invalid for extreme orbits with a small semi-major axis and a high value of
eccentricity (see \S \ref{orbital_configuration}).

\begin{figure}
  \includegraphics[angle=270,width=8.2cm]{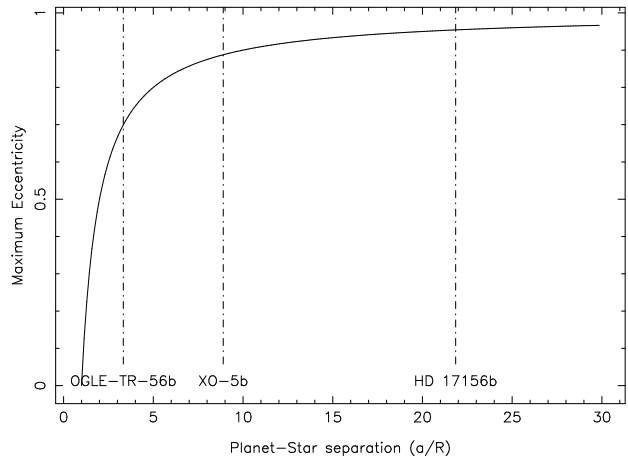}
  \caption{The maximum orbital eccentricity, $e_{\mathrm{max}}$,
    plotted as a function of the planet-star separation in units of
    $a/R$ (see Equation \ref{emax}) in order for a planet to remain
    outside the surface of its parent star. This requirement is purely
    based on system geometry and does not take into account tidal
    effects or planet-planet interactions, but requires that $a > R$
    when $f = 0$. Dot-dashed lines indicate values for $a/R$ for
    various transiting planets.}
  \label{fig5}
\end{figure}

To quantify this assumption, we use Equation \ref{separation} to
calculate the maximum eccentricity, $e_{\mathrm{max}}$, allowed as a
function of the planet-star separation in units of $a/R$ where $R
\equiv R_p + R_\star$. Applying the constraint $r > R$ when $f = 0$
(i.e., the planet is outside the star at periapsis) to Equation
\ref{separation} results in
\begin{equation}
  R = \frac{a (1 - e_{\mathrm{max}}^2)}{1 + e_{\mathrm{max}}}
\end{equation}
and thus,
\begin{equation}
  e_{\mathrm{max}} = 1 - \frac{R}{a}.
  \label{emax}
\end{equation}
Equation \ref{emax} is plotted in Figure \ref{fig5} for values of
$a/R$ ranging from 1 to 30. Also shown are dot-dashed lines which
indicate the $a/R$ values for OGLE-TR-56b \citep{kon03}, XO-5b
\citep{bur08b}, and HD 17156b.

The restrictions upon the maximum eccentricity begin to become
significant for $a/R < 10$, which encompasses most of the known
transiting exoplanets. This restriction is purely based upon orbital
dynamics and there are undoubtedly additional limitations on the
eccentricity due to tidal effects in this region.

\subsection{Orbital Inclination and Argument of Periastron}
\label{i_and_w}

One of the primary advantages of observing an exoplanet transiting the
host star is that it eliminates the ambiguity in the planetary mass
created by the unknown orbital inclination angle, $i$. The precise value of
the inclination can be derived from the impact parameter of the
transit across the stellar disk, defined by
\begin{equation}
  b \equiv \frac{a \cos i}{R_\star}
\end{equation}
and measurable from the shape of the lightcurve and the planet-star radius
ratio \citep{sea03}. Due to the constraint placed on $i$ by the presence of
transits, the true planetary mass will be within a few percent of the value
as determined from RV measurements alone.

The data available for transiting planets from the Extra-solar Planets
Encyclopaedia\footnote{http://exoplanet.eu/} and from \citet*{tor08}
show that the current distribution of inclination angles extends from
90\degr to almost 78\degr. The transiting planets whose orbits feature
numerically lower values of $i$ (i.e., more ``face-on'') are dominated
by the very hot Jupiters, such as OGLE-TR-56b which has an inclination
of $78.8\degr \pm 0.5\degr$ \citep{pon07}.

\begin{figure}
  \includegraphics[angle=270,width=8.2cm]{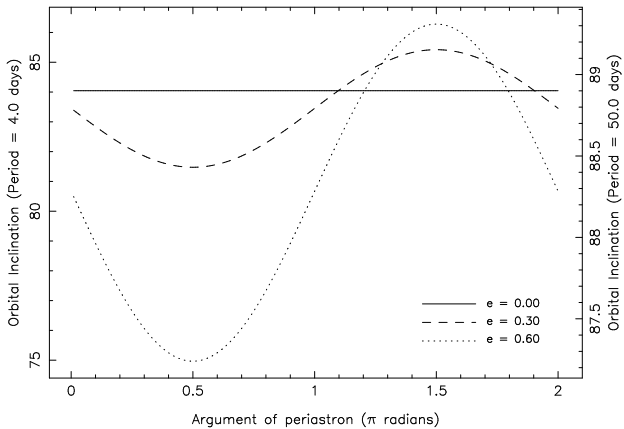}
  \caption{The maximum orbital inclination for a non-transiting planet
    as a function of the argument of periastron, $\omega$, for
    eccentricities of 0.0 (solid line), 0.3 (dashed line), and 0.6
    (dotted line), plotted for periods of both 4.0 days and 50.0
    days.}
  \label{fig6}
\end{figure}

If, however, a planet is determined not to transit, then limits may be
placed upon the orbital inclination if the eccentricity and argument
of periastron are known from RV measurements. This results from
re-expressing Equation \ref{transit_prob1} as follows:
\begin{equation}
  \cos i = \frac{R_p + R_\star}{a (1 - e \cos E)}.
  \label{cos_i}
\end{equation}
Figure \ref{fig6} shows the maximum inclination for various values of
period, $e$, and $\omega$. These are calculated by holding period and
$e$ fixed whilst varying $\omega$ using Equations \ref{cose} and
\ref{cos_i}. We further assume a Jupiter radius and a solar radius for
the values of $R_p$ and $R_\star$ respectively. For non-transiting
planets on orbits with $e \neq 0$ whose periastron is aligned towards
the observer (i.e., $\omega \sim \pi/2$), Figure \ref{fig6} shows that
the constraint on the inclination can be as high as $i \leq 75\degr$,
depending upon the orbital period. This is particularly useful for
those planets whose mass estimate places them close to the brown dwarf
regime. Note that Equation \ref{cos_i} reduces to Equation \ref{emax}
when $i = 90\degr$ and $\omega = \pi/2$, consistent with the
requirement that the planet remain outside the star.


\subsection{Orbital Inclination and Eccentricity}
\label{i_and_e}

As we show in \S \ref{i_and_w}, the fact that a planet is found to not
transit limits the possible combinations of $e$, $\omega$, and $i$. We
now consider what constraints may be placed upon the orbital
inclination for a non-transiting RV planet as a function of
eccentricity for the specific examples of when the periapsis is
aligned towards ($\omega \sim \pi/2$) and away from ($\omega \sim
3\pi/2$) the observer.

\begin{figure}
  \includegraphics[width=8.2cm]{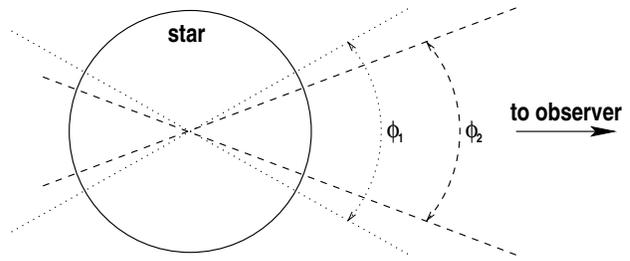}
  \caption{An edge-on view of two planetary orbits with the same
    values of semi-major axis, showing the range of excluded
    inclinations for an orbit with low eccentricity (dotted lines;
    $\phi_1$) and high eccentricity (dashed lines; $\phi_2$) for which
    the planet does not transit the parent star. For $\omega =
    3\pi/2$, the range of possible inclinations decreases with
    increasing eccentricity.}  \label{fig7}
\end{figure}

Figure \ref{fig7} illustrates the range of orbital inclinations that
are excluded for two orbits (shown edge-on) of non-transiting planets.
Both of these orbits have the same semi-major axes but different
eccentricities and are aligned such that $\omega = 3\pi/2$ (i.e.,
periapsis occurs behind the star as seen from the observer). In this
case, the range of possible values of $i$ increases with decreasing
orbital eccentricity ($\phi_1 > \phi_2$). The opposite is true when
$\omega = \pi/2$. In fact, the inclination in that case is only
constrained by the requirement that the planet remain outside the star
during periapsis (Equation \ref{emax}).

\begin{figure*}
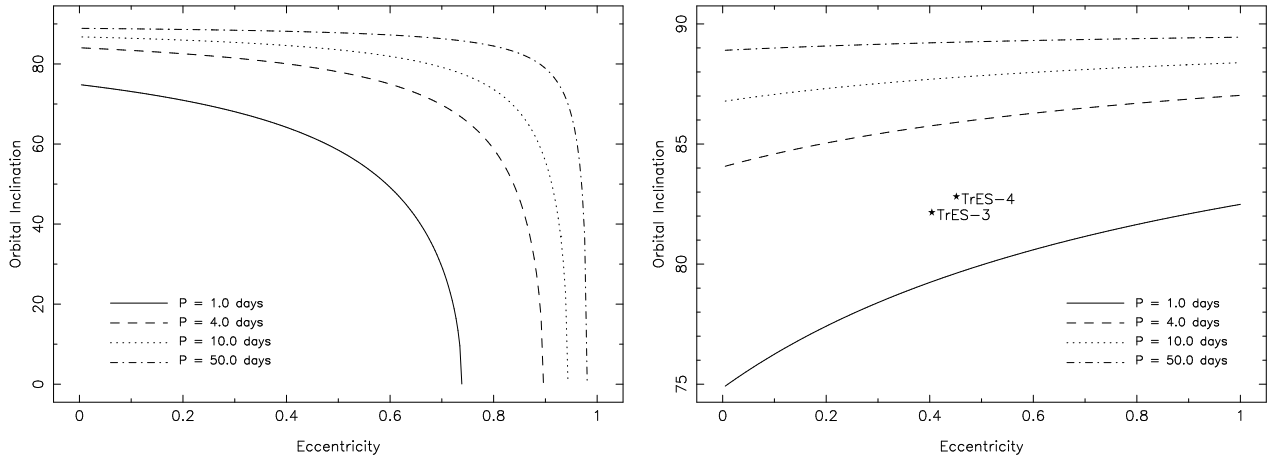

  \begin{center}
    \begin{tabular}{cc}
      \includegraphics[angle=270,width=8.2cm]{f8a.eps} &
      \includegraphics[angle=270,width=8.2cm]{f8b.eps} \\
    \end{tabular}
  \end{center}
  \caption{Maximum orbital inclination as a function of $e$ for
    non-transiting planets, plotted for four different periods (see
    Equation \ref{cos_i}). The left panel, case (a), is for $\omega =
    \pi/2$ and is based on the requirement that the planet remain
    outside the star (\S \ref{radii}).  The right panel, case (b),
    shows the situation for $\omega = 3\pi/2$, and is based on the
    geometrical arguments outlined in \S \ref{i_and_e} and shown in
    Figure \ref{fig7}. The location of the maximum eccentricities of
    the known transiting planets TrES-3 and TrES-4, given their
    parameters in Table \ref{tab1} and assuming $\omega = 3\pi/2$, are
    indicated by 5-pointed stars.}  \label{fig8}
\end{figure*}

Figure \ref{fig8} graphically demonstrates these constraints by plotting
Equation \ref{cos_i}, except now we fix the period and $\omega$ and
vary $e$. A Jupiter radius and a solar radius are assumed for the
values of $R_p$ and $R_\star$ respectively. The lines in these plots
represent the maximum values for $i$ for a non-transiting planet as a
function of $e$. These calculations are performed for four different
periods and the two aforementioned orientations of $\omega$: $\omega =
\pi/2$; left panel; case (a), and $\omega = 3\pi/2$; right panel; case
(b).  It is worth noting that case (a) in Figure \ref{fig8} represents
the physical constraint described in Figure \ref{fig5}, since Equation
\ref{cos_i} reduces to Equation \ref{emax} for $\omega = \pi/2$, as
stated above.

In case (a), for example, a non-transiting planet in a 4-day orbit
with $e=0.4$ has a range of possible inclination angles of $i \leq
80^{\degr}$. For larger values of $e$ (at any period), the periastron
distance of the planet will become so small that almost all values of
$i$ are possible, the maximum value of $e$ at each period being
defined by Equation \ref{emax}. The dependence of $i$ upon $e$ is
weaker for case (b) (note different scale for the left panel in Figure
\ref{fig8}) since the periastron passage now happens behind the star
(Figure \ref{fig2}) as seen from the observer, and thus, the range of
possible $i$-values is not very constrained by $e$.

We now consider a planet discovered using the transit method with
known $i$, $a$, $R_p$, and $R_\star$, but unknown values for $e$ and
$\omega$.  Is it possible to constrain $e$ in this scenario?  Using
Equation \ref{cos_i}, the eccentricity can be expressed as follows
\begin{equation}
  e = \frac{1}{cos E} \left( 1 - \frac{R_p + R_\star}{a \cos i}
  \right).
  \label{eccentricity}  
\end{equation}
However, there exists a degeneracy between $e$ and $\omega = f(E)$
such that one cannot place constraints on one parameter without
knowledge of the other. Additionally, as shown in \S \ref{i_and_w},
the constraint upon $i$ is only limited by the orbital boundary
defined by Equation \ref{emax} when $\omega = \pi/2$.

Therefore, a meaningful constraint may only be placed upon $e$ for
values of $\omega$ for which the orbital inclination is greater than
the maximum predicted for a circular orbit (i.e., the region above the
solid line shown in Figure \ref{fig6}). For case (b) ($\omega =
3\pi/2$), Equation \ref{eccentricity} reduces to
\begin{equation}
  e = \frac{R_p + R_\star}{a \cos i} - 1.
\end{equation}
As an example, consider the two known transiting planets TrES-3 and
TrES-4. The fit parameters shown in Table \ref{tab1} for the values of
$R_\star$, $R_p$, $i$, and $a$ are those reported by the discovery
papers for TrES-3 \citep{odo07} and TrES-4 \citep{man07}. Also shown
in Table \ref{tab1} are the maximum eccentricities for both case (a)
and case (b). For case (a), the maximum eccentricity is $\sim 0.8$ for
both planets.  For case (b), the maximum eccentricity for these two
planets is 0.4--0.5 and are plotted in the right panel of Figure
\ref{fig8}. In each case, the maximum eccentricities are remarkably
similar because the longer period of TrES-4 is compensated by the
relatively large radii of the star and planet.

\begin{table}
\begin{center}
  \caption{Fit parameters for TrES-3 and TrES-4 from \citet{odo07} and
    \citet{man07} respectively, along with the calculated maximum
    eccentricities.}
  \label{tab1}
  \begin{tabular}{@{}ccc}
    \hline
    Parameter & TrES-3 & TrES-4 \\
    \hline
    $R_\star$          & $0.802 \pm 0.046$   & $1.738 \pm 0.092$   \\
    $R_p$              & $1.295 \pm 0.081$   & $1.674 \pm 0.094$   \\
    $i$                & $82.15 \pm 0.21$    & $82.81 \pm 0.33$    \\
    $a$                & $0.0226 \pm 0.0013$ & $0.0488 \pm 0.0022$ \\
    $e(\omega=\pi/2)$  & $0.808 \pm 0.083$   & $0.818 \pm 0.073$   \\
    $e(\omega=3\pi/2)$ & $0.404 \pm 0.041$   & $0.451 \pm 0.040$   \\
    \hline
  \end{tabular}
\end{center}
\end{table}


\subsection{Global Statistics}

The total number of transiting planets discovered thus far via radial
velocity surveys does not necessarily reflect the true number of
transiting planets in this sample. At the time of writing, most of the
known radial velocity planets have not been adequately monitored
photometrically in order to rule out transits. We can estimate the
number of planets that should be transiting, and determine the
significance of a hypothetical null result from a photometric
follow-up campaign by applying the results of this paper to the
\citet{but06} RV planet sample.

\begin{figure}
  \includegraphics[angle=270,width=8.2cm]{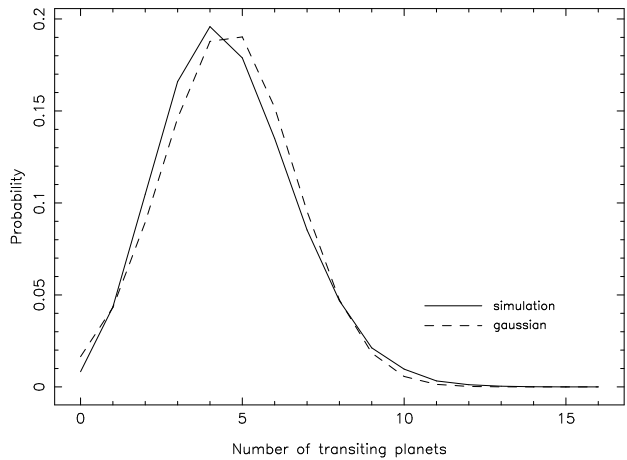}
  \caption{The probability distribution (solid line) for the 203
    planets in the \citet{but06} sample, predicting the number of
    transiting planets based on their estimated orbital
    parameters. Over-plotted is a gaussian distribution (dashed line)
    using the mean and standard deviation of the simulation
    results.}
  \label{fig9}
\end{figure}

The host star properties and planetary orbital parameters provided by
\citet{but06} form the foundation of a Monte-Carlo simulation of the
transit probabilities calculated from Equation
\ref{transit_prob2}. The planetary radii $R_p$ are assumed to be one
Jupiter radius as used in previous sections. However, the stellar
radii $R_\star$ are estimated individually from the values of $B - V$
provided by \citet{but06}, assuming the host stars are dwarf stars
\citep{cox00}. The orbital elements $a$, $e$ and $\omega$ are directly
extracted from \citet{but06}. Using these values, we calculate $P_t$
for each of the 203 stars in the sample and randomly determine if the
planet transits. This yields an integer number of projected transits
from the sample. By performing these calculations $\sim 100000$ times,
we produce a probability distribution for the number of transiting
planets expected from this sample, shown in Figure \ref{fig9}.

The simulated probability distribution has a mean value of $\sim 4.5$
transits peaking at $P_t \sim 0.2$ with a standard deviation of $\sim
2.0$. For comparison, we also generated a gaussian distribution
profile using this mean and standard deviation. The a priori
probability that none of the planets in this sample transit their host
star is $\sim 1$\%. In fact, three of the planets in this sample are
known to transit, specifically HD 17156b, GJ 436b, and HD
147506b. Hence the current number of transiting planets from this
sample is almost 1$\sigma$ below the expectation.

We further note that the sample of RV planets is biased toward
numerically higher values of $i$ since detection efficiency will
increase with higher $i$ for a given RV precision. As such, the
expected number of transiting planets in the sample should be regarded
as a lower limit. Though the discrepancy between known and expected
transiting planets is not significant in this low-number regime, it is
nevertheless quantifiable, and we conclude that further transit
discoveries in this sample are possible or even likely.  Any such
additional detections would, in turn, lead to further understanding of
the respective observational biases of the RV and transit methods. For
example, the observational bias leads to an observed difference
between the period distributions of planets discovered by the transit
method and the radial velocity method, as discussed in detail by
\citet*{gau05}.


\section{Conclusions}

It is still uncertain at this stage how many of the known radial
velocity planets transit their parent stars. What is clear is that the
eccentricity distribution of the known exoplanets will increase the
transit likelihood, making detections for long-period planets, such as
HD 17156b, feasible. We have shown in this paper that there is enough
potential amongst longer period planets for transit detections to
motivate a photometric monitoring campaign at the predicted times of
transit for these targets. \citet{fle08} have shown that long-period
transiting planets may yet be discovered through ground-based transit
surveys, particularly if data sets from different surveys are
combined.

As pointed out by \citet{bar07b}, eccentric planets that have a
periastron oriented away from the observer are far more likely to
exhibit a secondary than a primary eclipse. The detection of such a
secondary eclipse is considerably more challenging than for a primary
eclipse since it relies on a minimum level of planetary flux and is
best pursued at infrared wavelengths. The discussion in \S
\ref{i_and_e} shows that even an assumption of $\omega = 3\pi/2$ can
place constraints on the orbital inclination. A prime candidate for
such a study is HD 80606b \citep{nae01} which has a period of 111.87
days and an eccentricity of 0.927. Scaling Figure \ref{fig1} to this
period and eccentricity yields a secondary transit probability of
$\sim 15$\%.

Many of the results presented in this paper can easily be applied to
any system since the results generally scale linearly with the sum of
the stellar and planetary radii. Through applying these results to
current and future radial velocity planet discoveries, one can choose
targets for an efficient observing campaign which may help to discover
long-period transiting planets and hence add invaluable information to
planetary structure and formation theories.


\section*{Acknowledgements}

The authors would like to thank David Ciardi, Scott Fleming, and Alan
Payne for several useful discussions. We would especially like to
thank the referee Jason W. Barnes, who provided a fast and insightful
report which greatly improved the quality of the paper.




\begin{thebibliography}{}

\bibitem[\protect\citeauthoryear{Anderson et al.}{2008}]{and08}
  Anderson, D.R., et al., 2008, MNRAS, 387, L4
\bibitem[\protect\citeauthoryear{Armitage}{2007}]{arm07} Armitage,
  P.J., 2007, ApJ, 665, 1381
\bibitem[\protect\citeauthoryear{Barbieri et al.}{2007}]{bar07a}
  Barbieri, M., et al., 2007, A\&A, 476, L13
\bibitem[\protect\citeauthoryear{Barnes}{2007}]{bar07b} Barnes, J.W.,
  2007, PASP, 119, 986
\bibitem[\protect\citeauthoryear{Bouchy et al.}{2005}]{bou05} Bouchy,
  F., et al., 2005, A\&A, 444, L15
\bibitem[\protect\citeauthoryear{Borucki \& Summers}{1984}]{bor84}
  Borucki, W.J., Summers, A.L., 1984, Icarus, 58, 121
\bibitem[\protect\citeauthoryear{Burke}{2008}]{bur08a} Burke, C.J.,
  2008, ApJ, 679, 1566
\bibitem[\protect\citeauthoryear{Burke}{2008}]{bur08b} Burke, C.J.,
  et al., 2008, ApJ, submitted (arXiv:0805.2399)
\bibitem[\protect\citeauthoryear{Butler et al.}{2006}]{but06} Butler,
  R.P., et al., 2006, ApJ, 646, 505
\bibitem[\protect\citeauthoryear{Charbonneau et al.}{2000}]{cha00}
  Charbonneau, D., Brown, T.M., Latham, D.W., Mayor, M., 2000, ApJ,
  529, L45
\bibitem[\protect\citeauthoryear{Cox}{2000}]{cox00} Cox, A.N., 2000,
  Allen’s Astrophysical Quantities (4th ed.; New York: AIP)
\bibitem[\protect\citeauthoryear{Fleming et al.}{2008}]{fle08}
  Fleming, S.W., Kane, S.R., McCullough, P.R., Chromey, F.R., 2008,
  MNRAS, 386, 1503
\bibitem[\protect\citeauthoryear{Ford \& Rasio}{2008}]{for08a} Ford,
  E.B., Rasio, F.A., 2008, ApJ, in press (astro-ph/0703163)
\bibitem[\protect\citeauthoryear{Ford, Quinn, \& Veras}{Ford et
    al.}{2008}]{for08b} Ford, E.B., Quinn, S.N., Veras, D., 2008,
  ApJ, 678, 1407
\bibitem[\protect\citeauthoryear{Gaudi, Seager, \&
    Mallen-Ornelas}{Gaudi et al.}{2005}]{gau05} Gaudi, B.S., Seager,
  S., Mallen-Ornelas, G., 2005, ApJ, 623, 472
\bibitem[\protect\citeauthoryear{Gillon et al.}{2007}]{gil07} Gillon,
  M., et al., 2007, A\&A, 472, L13
\bibitem[\protect\citeauthoryear{Henry et al.}{2000}]{hen00} Henry,
  G.W., Marcy, G.W., Butler, R.P., Vogt, S.S., 2000, ApJ, 529, L41
\bibitem[\protect\citeauthoryear{Johns-Krull et al.}{2008}]{joh08}
  Johns-Krull, C.M., et al., 2008, ApJ, 677, 657
\bibitem[\protect\citeauthoryear{Kane}{2007}]{kan07a} Kane, S.R.,
  2007, MNRAS, 380, 1488
\bibitem[\protect\citeauthoryear{Kane, Schneider, \& Ge}{Kane et
    al.}{2007}]{kan07b} Kane, S.R., Schneider, D.P., Ge, J., 2007,
  MNRAS, 377, 1610
\bibitem[\protect\citeauthoryear{Konacki et al.}{2003}]{kon03}
  Konacki, M., Torres, G., Jha, S., Sasselov, D.D., 2003, Nature, 421,
  507
\bibitem[\protect\citeauthoryear{Li et al.}{2008}]{li08} Li,
  C.-H., et al., 2008, Nature, 452, 610
\bibitem[\protect\citeauthoryear{L\'opez-Morales}{2006}]{lop06a}
  L\'opez-Morales, M., 2006, PASP, 118, 716
\bibitem[\protect\citeauthoryear{L\'opez-Morales et al.}{2006}]{lop06b}
  L\'opez-Morales, M., Morrell, N.I., Butler, R.P., Seager, S., 2006,
  PASP, 118, 1506
\bibitem[\protect\citeauthoryear{Mandushev et al.}{2007}]{man07}
  Mandushev, G., et al., 2007, ApJ, 667, L195
\bibitem[\protect\citeauthoryear{Marcy et al.}{1997}]{mar97} Marcy,
  G.W., Butler, R.P., Williams, E., Bildsten, L., Graham, J.R., Ghez,
  A.M., Jernigan, G., 1997, ApJ, 481, 926
\bibitem[\protect\citeauthoryear{Naef et al.}{2001}]{nae01} Naef, D.,
  et al., 2001, A\&A, 375, L27
\bibitem[\protect\citeauthoryear{O'Donovan et al.}{2007}]{odo07}
  O'Donovan, F.T., et al., 2007, ApJ, 663, L37
\bibitem[\protect\citeauthoryear{P\'al et al.}{2008}]{pal08}
  P\'al, A., et al., 2008, ApJ, in press (arXiv:0803.0746)
\bibitem[\protect\citeauthoryear{Pepe et al.}{2004}]{pep04} Pepe, F.,
  et al., 2004, A\&A, 423, 385
\bibitem[\protect\citeauthoryear{Pont et al.}{2007}]{pon07}
  Pont, F., et al., 2007, A\&A, 465, 1069
\bibitem[\protect\citeauthoryear{Sato et al.}{2005}]{sat05} Sato, B.,
  et al., 2005, ApJ, 633, 465
\bibitem[\protect\citeauthoryear{Seager \&
    Mall\'en-Ornelas}{2003}]{sea03} Seager, S., Mall\'en-Ornelas, G.,
  2003, ApJ, 585, 1038
\bibitem[\protect\citeauthoryear{Shankland et al.}{2006}]{sha06}
  Shankland, P.D., et al., 2006, ApJ, 653, 700
\bibitem[\protect\citeauthoryear{Torres, Winn, \& Holman}{Torres et
    al.}{2008}]{tor08} Torres, G., Winn, J.N., Holman, M.J., 2008,
  ApJ, 677, 1324
\bibitem[\protect\citeauthoryear{Winn et al.}{2007}]{win07} Winn,
  J.N., et al., 2007, ApJ, 665, L167
\end{thebibliography}
\end{document}